\documentstyle{mn}     
\input{epsf}    
\newif\ifAMStwofonts    
\def\simlt{\lower.5ex\hbox{$\; \buildrel < \over \sim \;$}}   
\def\simgt{\lower.5ex\hbox{$\; \buildrel > \over \sim \;$}}  
\def\eps@scaling{.95}  
\def\epsscale#1{\gdef\eps@scaling{#1}}  
  
\def\plotone#1{\centering \leavevmode  
    \epsfxsize=\eps@scaling\columnwidth \epsfbox{#1}}

\def\lbd{$\Lambda$}   

\newif\ifAMStwofonts

\ifoldfss       
  \ifCUPmtlplainloaded \else       
    \NewTextAlphabet{textbfit} {cmbxti10} {}       
    \NewTextAlphabet{textbfss} {cmssbx10} {}       
    \NewMathAlphabet{mathbfit} {cmbxti10} {} 
    \NewMathAlphabet{mathbfss} {cmssbx10} {} 
  \fi       
  \ifAMStwofonts       
    \ifCUPmtlplainloaded \else       
      \NewSymbolFont{upmath} {eurm10}       
      \NewSymbolFont{AMSa} {msam10}       
      \NewMathSymbol{\upi}     {0}{upmath}{19}       
      \NewMathSymbol{\umu}     {0}{upmath}{16}       
      \NewMathSymbol{\upartial}{0}{upmath}{40}       
      \NewMathSymbol{\leqslant}{3}{AMSa}{36}       
      \NewMathSymbol{\geqslant}{3}{AMSa}{3E}

    \fi       
  \fi       
\fi 
       
\ifnfssone       
  \newmathalphabet{\mathit}       
  \addtoversion{normal}{\mathit}{cmr}{m}{it}       
  \addtoversion{bold}{\mathit}{cmr}{bx}{it}       
  \newmathalphabet{\mathbfit} 
  \addtoversion{normal}{\mathbfit}{cmr}{bx}{it}       
  \addtoversion{bold}{\mathbfit}{cmr}{bx}{it}       
  \newmathalphabet{\mathbfss} 
  \addtoversion{normal}{\mathbfss}{cmss}{bx}{n}       
  \addtoversion{bold}{\mathbfss}{cmss}{bx}{n}       
  \ifAMStwofonts       
    \ifCUPmtlplainloaded \else       
      \UseAMStwoboldmath       
      \makeatletter       
      \new@mathgroup\upmath@group       
      \define@mathgroup\mv@normal\upmath@group{eur}{m}{n}       
      \define@mathgroup\mv@bold\upmath@group{eur}{b}{n}       
      \edef\UPM{\hexnumber\upmath@group}       
      \new@mathgroup\amsa@group       
      \define@mathgroup\mv@normal\amsa@group{msa}{m}{n}       
      \define@mathgroup\mv@bold\amsa@group{msa}{m}{n}       
      \edef\AMSa{\hexnumber\amsa@group}       
      \makeatother       
      \mathchardef\upi="0\UPM19       
      \mathchardef\umu="0\UPM16       
      \mathchardef\upartial="0\UPM40       
      \mathchardef\leqslant="3\AMSa36       
      \mathchardef\geqslant="3\AMSa3E       
    \fi       
  \fi       
\fi 
       
\ifnfsstwo       
  \DeclareMathAlphabet{\mathbfit}{OT1}{cmr}{bx}{it}       
  \SetMathAlphabet\mathbfit{bold}{OT1}{cmr}{bx}{it}       
  \DeclareMathAlphabet{\mathbfss}{OT1}{cmss}{bx}{n}       
  \SetMathAlphabet\mathbfss{bold}{OT1}{cmss}{bx}{n}       
  \ifAMStwofonts       
    \ifCUPmtlplainloaded \else       
      \DeclareSymbolFont{UPM}{U}{eur}{m}{n}       
      \SetSymbolFont{UPM}{bold}{U}{eur}{b}{n}       
      \DeclareSymbolFont{AMSa}{U}{msa}{m}{n}       
      \DeclareMathSymbol{\upi}{0}{UPM}{"19}       
      \DeclareMathSymbol{\umu}{0}{UPM}{"16}       
      \DeclareMathSymbol{\upartial}{0}{UPM}{"40}       
      \DeclareMathSymbol{\leqslant}{3}{AMSa}{"36}       
      \DeclareMathSymbol{\geqslant}{3}{AMSa}{"3E}       
    \fi       
  \fi       
\fi 
       
\ifCUPmtlplainloaded \else       
  \ifAMStwofonts \else 
    \def\upi{\pi}       
    \def\umu{\mu}       
    \def\upartial{\partial}       
  \fi       
\fi

\newcommand{\Msun}{\>{\rm M_{\odot}}}   
   
\newcommand{\mpch}{\>h^{-1}{\rm {Mpc}}}

\newcommand{\Msunh}{\>h^{-1}\rm M_\odot}   
\def \etal {et~al.~}   
\def\apj{ApJ}
\def\mnras{MNRAS}

\def\aap{A\&A}

\title[Constraining WDM using QSO gravitational lensing ]   
{Constraining Warm Dark Matter using QSO gravitational lensing }       
   
\author[M. Miranda \& A.V. Macci\`o]      
{Marco Miranda$^1$\thanks{E-mail: solar@physik.unizh.ch} and Andrea V. Macci\`o$^2$\thanks{E-mail: maccio@mpia.de}  \\      
$^1$Institute for Theoretical Physics, University of Z\"urich,        
Winterthurerstrasse 190, CH-8057 Z\"urich, Switzerland \\       
$^2$Max-Planck-Institute for Astronomy, K\"onigstuhl 17,   
           D-69117 Heidelberg, Germany\\   
}   
   
\smallskip        
       
\pubyear{2007}       
       
\begin{document}

\maketitle       
       
\begin{abstract}

Warm Dark Matter (WDM) has been invoked to resolve apparent conflicts of Cold Dark 
Matter (CDM) models with observations on subgalactic scales. 
In this work we provide a new and independent lower limit for the WDM particle mass 
(e.g. sterile neutrino) through the analysis of image fluxes in gravitationally lensed 
QSOs.  

Starting from a theoretical unperturbed cusp configuration we analyze the effects of 
intergalactic haloes in modifying 
the fluxes of QSO multiple images, giving rise to the so-called anomalous flux ratio. 
We found that the global effect of such haloes strongly depends 
on their mass/abundance ratio and it is maximized for haloes in the mass range 
$10^6-10^8 \Msun$. 
  
This result opens up a new possibility to constrain CDM predictions on small scales and 
test different warm candidates, since free streaming of warm dark matter particles 
can considerably dampen the matter power spectrum in this mass range. 
As a consequence, while a ($\Lambda$)CDM model is able to produce 
flux anomalies at a level similar to those observed, a WDM model, with an
insufficiently massive particle, fails to reproduce the observational evidences. 

Our analysis suggests a lower limit of a few keV ($m_{\nu} \sim 10$) 
for the mass of warm dark matter candidates in the form of a sterile neutrino.
This result makes sterile neutrino Warm Dark Matter less attractive 
as an alternative to Cold Dark Matter, in good agreement with previous findings from 
Lyman-$\alpha$ forest and Cosmic Microwave Background analysis.

\end{abstract}


\section{Introduction}       
   
The Cold Dark Matter (CDM) model  
has been successful in explaining a large variety of observational results      
such as the large scale structure of the universe and 
fluctuations of the Cosmic Microwave Background (CMB, Spergel \etal 2003, 2006).   
However, the CDM model faces some apparent problems on small scales: namely the overprediction   
of galactic satellites, the cuspiness and high density of galactic cores and the large number  
of galaxies filling voids (Klypin \etal 1999, Moore \etal 1999a,b, Bode, Ostriker \& Turok 2001,  
Avila-Reese \etal 2001, Peebles 2001 and references therein).   
These problems may well have complex astrophysical solutions.  
For instance the excess of galactic satellites can be alleviated by 
feedback processes such as heating and supernova winds that can inhibit the star formation   
in low-mass haloes (Bullock, Kravtsov \& Weinberg 2001).
   
Another natural cosmological solution to these problems is to replace cold   
dark matter with a warm species (\lbd WDM, see Bode, Ostriker \& Turok 2001   
and references therein). The warm component acts to reduce   
the small-scale power, resulting in fewer galactic subhaloes and lower central
halo densities.  
   
One of the most promising WDM candidates is a sterile (right-handed) neutrino with a mass in  
the keV range; such a particle may occur naturally  within extensions to the standard model of  
particle physics (Dodelson \& Widrow 1994, Dolgov \& Hansen 2002, Asaka \etal 2005, 
Viel \etal  2005). A sterile neutrino is non-thermal in extensions of the  
minimal standard model, with a life-time longer than the age of the universe.   
   
A strong constraint on the mass of WDM candidates comes from Lyman-$\alpha$ forest observations  
(neutral hydrogen absorption in the spectra of distant quasars), since they are a powerful  
tool for constraining the matter power spectrum over a large range of redshifts down to small  
scales. Recent analysis of SDSS quasar spectra combined with CMB and galaxy
clustering data have set a lower limit on the mass of the sterile neutrino around $m_{\nu} \approx 10-13$  
keV (Seljak \etal  2006, Viel \etal 2006).   
In this paper we use a completely different approach to put independent constraints on $m_{\nu}$,  
using QSO gravitational lensing and the so-called anomalous flux ratio.  

Standard lens models, although they reproduce in general the relative positions     
of the images quite accurately, often have difficulties explaining the      
relative fluxes of multiply-imaged sources (Mao \& Schneider 1998, Metcalf \&
Madau 2001, Dalal \& Kochanek 2002, Metcalf and Zhao 2002), giving rise to the  
so-called anomalous flux ratio problem.  
   
Several possible explanations have been considered in the literature, the most     
plausible being that the lensing potential of real galaxies are not     
fully described by the simple lens models used to compute lens     
characteristics.
The most often invoked solution is to consider additional small-scale     
perturbations (i.e. dark matter haloes), which if located near a photon's light  
path can modify the overall lens potential (e.g. Raychaudhury \etal 2000, Saha \etal 2007) and 
significantly alter the observed flux ratio between different images,
in particular in the cusp or fold configuration (Metcalf \& Madau 2001,
Chiba 2002, Chen \etal 2003, Metcalf 2005a,b, Dobler \& Keeton 2006).
Those perturbers can be roughly divided in two categories: haloes that are inside the primary lens,  
usually referred as sub-haloes, and haloes that are along the line of sight, in between the  
source and the observer.
This first category of haloes has been extensively studied in the past years both through analytic  
calculation (Metcalf \& Madau 2001, Dalal \& Kochanek 2002, Metcalf and Zhao 2002, Keeton 2003) 
and using numerical  
simulations (Brada\v c \etal 2002, Amara \etal 2006, Macci\`o \etal 2006).   
The latter two studies have came to the conclusion that the impact of sub-haloes on lensing in  
the mass range $10^7-10^{10} \Msunh$ is very small. Even considering the impact of less massive   
subhaloes, usually not resolved in Nbody/hydro simulations, does not help in reproducing the  
observed number of anomalous flux ratios (Macci\`o \& Miranda 2006).  
   
The effect of the second category of haloes, those along the line of sight,  
is still somewhat controversial (Chen \etal 2003, Metcalf 2005a,b).  
In particular Metcalf (2005a,b) found that dark matter haloes with masses around  
$10^6-10^8 \Msun$ can produce anomalies in the flux ratios at a level similar to those  
that are observed.  
The presence of a WDM particle even with a mass around $10$ keV will strongly reduce  
the number density of such small mass haloes, giving a different signature to the  
image fluxes.   
As a consequence, the observed anomalous flux ratios can be used to constrain the abundance of  
small haloes along the line of sight and therefore to put an independent constraint on the  
mass of the sterile neutrino as a possible WDM candidate.   
   
In this paper we analyze in detail the effect of subhaloes along the line of sight   
on an unperturbed cusp configuration in a $\Lambda$CDM  model and in $\Lambda$WDM  
models with different values of $m_{\nu}$.
We found that WDM models with $m_{\nu}<10$ keV fail to reproduce the observed anomalies in  
the lensed QSO flux ratios. Our results provide a new and independent  
constraint on the mass of sterile neutrino, and they are in good agreement with 
previous constraints coming from Lyman-$\alpha$ forest and CMB analysis.   
   
The format of the paper is as follows:  
in section \ref{sec:haloes} we compute the expected halo abundance  
in different models; in section \ref{sec2los} we review briefly the lensing formalism we adopt.   
Section \ref{sec:sim} is devoted to the description of our lensing simulations.   
In section \ref{sec:res} we present the numerical results, matching them with observations.      
We conclude with a short summary and discussion of our results in section \ref{sec:disc}.

\section{Intergalactic halo mass function}       
\label{sec:haloes}   
   
The main goal of this work is to study the effect of dark matter haloes along the line of  
sight on fluxes of QSO multiple images. In order to achieve it we first computed
the number density of those haloes in the light cone between the source plane and the observer.  
   
For this purpose we used the Sheth and Tormen mass function (ST: Sheth \& Tormen 2002), 
taking into account its evolution with redshift.  
We adopted a WMAP1-like cosmology (Spergel \etal 2003) with the following values for  
dark energy and dark matter density, normalization and slope of the matter power spectrum:   
$\Omega_{\Lambda}=0.74$, $\Omega_{m}=0.26$, $\sigma_8=0.9$ and $n=1$.  
   
The transfer function for the CDM model has been generated using the public code CMBFAST   
(Seljak \& Zaldarriaga 1996). To compute the transfer function for WDM models we used the  
fitting formula suggested by Bode, Turok and Ostriker (2001):   
   
\begin{equation}   
T^2(k) = {P^{WDM} \over P^{CDM} } = [1+(\alpha k )^{2 \nu}]^{-10/\nu}   
\label{eq:Twdm}   
\end{equation}   
\noindent   
where $\alpha$, the scale of the break, is a function of the WDM parameters, while the index  
$\nu$ is fixed. Viel \etal (2005, see also Hansen \etal 2002), using a Boltzmann code
simulation, found that $\nu=1.12$ is the best fit for $k<5 ~h~ \rm Mpc^{-1}$,  
and they obtained the following expression for $\alpha$:   
\begin{equation}   
\alpha = 0.049  \left ( {m_x \over  \rm{1 keV}}  \right )^{-1.11}   \left ( { \Omega_{\nu}  
\over 0.25 }\right )^{0.11}   \left ( { h \over 0.7} \right )^{1.22} \mpch.   
\label{eq:alpha}
\end{equation}   
This expression applies only to the case of thermal relics.
In order to apply it to a sterile neutrino we take advantage of the one-to-one 
correspondence between the masses of 
thermal WDM particles ($m_x$) and sterile neutrinos ($m_\nu$) for which the effect on the 
matter distribution and thus the transfer function for both models are identical 
(Colombi \etal 1996).
We used the $m_x-m_{\nu}$ relation given by Viel \etal (2005), that reads:
\begin{equation}
m_{\nu,\rm{sterile}}
=  4.43 
\left({m_{x,\rm {thermal}} \over 1 \, \rm{keV}}\right)^{4/3}
\!
\left({0.25  \over \Omega_{\nu} } \right)^{1/3}  \left( {0.7 \over h } \right) ^{2/3}   \rm{keV}.
\end{equation}
We used the expression given in eq:\ref{eq:alpha} for the damping of the power-spectrum
for simplicity and generality. More accurate expressions for the damping for
concrete models of sterile neutrinos exist (Abazajian 2006, Asaka \etal 2007) 
and show that the damping depends on the detailed physics of the early universe
in a rather non-trivial way. Naturally the results of this paper can be
repeated using other expressions for the damping.

\begin{figure}   
\plotone{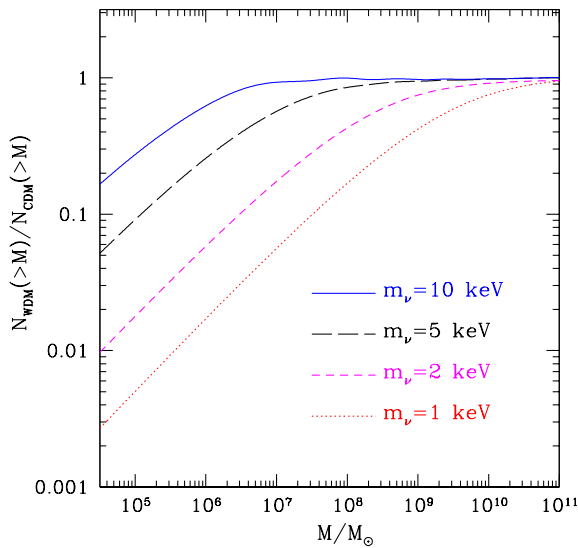}   
\caption{Effects of WDM particles on the dark matter halo mass function at redshift zero.}
\label{fig:mf}   
\end{figure}

The main effect of WDM is to dampen the power spectrum of fluctuation on small scales,  
reducing the number of haloes at low masses (Bode, Turok \& Ostriker 2001, Barkana \etal 2001, 
Paduroiu \etal 2007 in prep.).   
Figure \ref{fig:mf} shows the ration between halo number density in WDM and CDM models as  
a function of the WDM mass $m_{\nu}$.   

\begin{figure}   
\plotone{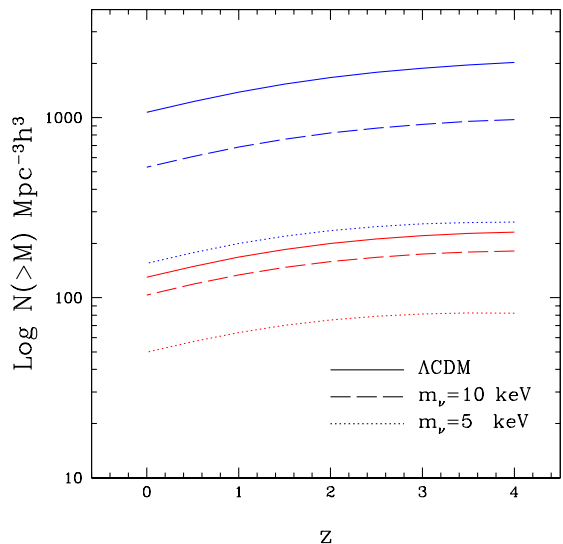}
\caption{Evolution with redshift of the number of haloes above a fixed mass threshold in
different models. The upper-most solid (blue) line is for $M>10^6 \Msunh$ in the $\Lambda$CDM model;
the dashed and the dotted lines are for the same mass threshold but for a WDM mass
of $m_{\nu}=10,5 \rm keV$ respectively.
The second set of (red) lines refers to a mass threshold of $M>10^7 \Msunh$.}
\label{fig:mfz}   
\end{figure}

Typically lensed QSOs are located at a redshift around 3.  
This implies that we also need to take into account the redshift evolution of the mass  
function in different models. Figure \ref{fig:mfz} shows  
the number of haloes more massive than $10^6 \Msunh$ (upper solid curve) and $10^7 \Msunh$  
(lower solid curve) per Mpc cube at different redshifts.  
It is interesting to note that on such small mass scales the halo number density tends  
to increase towards high redshift.  
We found that the evolution of the mass function,both in CDM and WDM models, can be well  
represented by the following fitting formula:   
\begin{equation}  
\log N(>M,z) =  N_0 + 0.11 \cdot  z^{0.7}   
\label{eq:fit}   
\end{equation}   
where $N_0$ is the logarithm of the halo number density at redshift zero  
($N_0=\log N(>M,z=0)$). The use of this fitting formula has the advantage of speeding up the  
calculation of the number of haloes in each lensing plane (see section \ref{sec:sim}).   
   
To conclude this section we want to emphasize that our particular choice of cosmological  
parameters does not influence the results we will present in the next section.  
For instance on the mass scales we are interested in ($M<10^{10} \Msunh$)  
changing $\sigma_8$ from 0.9 to 0.7 would increase the number of haloes only by a few percent.  

\section{Lensing Formalism} 
\label{sec2los}   
We briefly recall the general expressions for       
gravitational lensing and refer, e.g., to the book by Schneider \etal (1992)       
for more details.      
The lens equation is defined as:        
\begin{equation}       
\label{eq:lens}       
\vec{\theta}  =\vec{\beta} + \vec\alpha(\vec\theta)~,      
\end{equation}       
where $\vec\beta(\vec\theta)$ is       
the source position and $\vec\theta$ the image position.       
$\vec\alpha(\vec\theta)$ is the deflection angle, which depends       
on $\kappa(\vec\theta)$ the dimensionless surface mass density (or      
convergence) in units of the critical surface       
mass density $\Sigma_{\rm crit}$, defined as:   
\begin{equation}       
\label{eq:crit}       
\Sigma_{\rm crit} = { c^2  \over {4 \pi G }} { D_S  \over {D_L D_{LS} }},       
\end{equation}       
where $D_S, D_L, D_{LS}$ are the angular diameter distances between       
observer and source, observer and lens, source and lens, respectively.

\subsection{The cusp relation}   
   
\begin{figure}   
\plotone{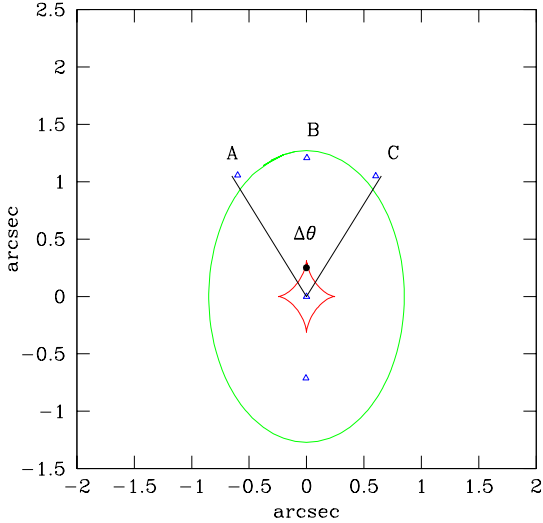}   
\caption{Unperturbed cusp configuration: $R_{cusp} = 0.09$. 
The source and image positions are marked  
by a solid circle and open triangles respectively. The opening angle is also shown.}   
\label{fig:conf1}   
\end{figure}   
There are basically three configurations of four-image systems: fold, cusp,   
and cross (Schneider \&  Weiss 1992). In this paper we will mainly  
concentrate on the {\it cusp}  
configuration, that corresponds to a source located close to the cusp of the   
inner caustic curve (see figure \ref{fig:conf1}). 
The behavior of gravitational lens mapping near a cusp   
was first studied by Blandford \& Narayan (1986), Schneider \& Weiss (1992),   
Mao (1992) and Zakharov (1995), who investigated the magnification 
properties of cusp images and concluded that the sum of the signed magnification 
factors of the three merging images approaches zero as the source moves 
towards the cusp. In other words:   
\begin{equation}   
R_{cusp} = {{ \mu_A + \mu_B + \mu_C} \over { \vert \mu_A \vert + \vert  
\mu_B \vert + \vert\mu_C \vert}} \rightarrow 0, ~~for ~~~~\mu_{tot} ~\rightarrow \infty   
\label{eq:rcusp}   
\end{equation}   
where $\mu_{tot}$ is the unsigned sum of magnifications of all four images,   
and A,B \& C are the triplet of images forming the smallest opening angle (see   
figure \ref{fig:conf1}).  
By opening angle, we mean the angle measured from the galaxy center and 
spanned by two images of equal parity. The third image lies inside such an angle.   
This is an asymptotic relation and holds when the source approaches   
the cusp from inside the inner caustic `` astroid''. 
This can be shown by expanding the lensing map to third order in the angular   
separation from a cusp (Schneider \&  Weiss 1992).  
Structure on scales smaller than the image separation   
will cause $R_{cusp}$ to differ from zero   
fairly independently of the form of the rest of the lens. 
Note that by definition of $R_{cusp}$ used here, it can be either
positive or negative. A perturber
is more likely to reduce the absolute magnification for negative   
magnification images (Metcalf \& Madau 2001, Schechter \&  Wambsganss 2002,   
Keeton et al. 2003) and to increase it for positive parity images. { As a
result, the probability  distribution of $R_{cusp}$ will be skewed toward
positive values.}
       
\subsection{The unperturbed lens}

We used the {\sc GRAVLENS} code (Keeton 2001)\footnote{The software is     
available via the web site http://cfa-www.harvard.edu/castles}   
to create a lens configuration for which the cusp  
relation is roughly satisfied (see figure \ref{fig:conf1}).   
The main, smooth, lens has been modelled as a singular isothermal ellipsoid   
(SIE) (Kormann, Schneider, \& Bartelmann 1994) to take advantage of its   
simplicity.  This model has been widely used in lens modeling and successfully   
reproduces many lens systems (e.g. Keeton \etal 1998, Chiba   
2002, Treu \& Koopmans 2002).  
The ellipsoidal primary lens has a mass equal to $5 \times 10^{11} \Msun$, is   
oriented with the major axis along the y axis in the lens plane and has an
ellipticity of 0.33. The redshift of the lens has been fixed to $z_l=0.3$ in   
agreement with typical observed ones (i.e. Tonry 1998).   
The cusp relation, defined by equation \ref{eq:rcusp}, for this smooth lens gives  
$R_{cusp}=0.09$, and this is one of the configurations previously studied 
in Macci\`o and Miranda (2006, namely Config2). 
We tested that our results do not 
depend on this particular choice for the unperturbed configuration and do apply 
to any cusp configuration.


\section{Subhaloes along the line of sight: Idea and Procedure}         
\label{sec:sim}

The purpose of this work is to compute the effects of intergalactic haloes, along the  
line of sight, on an unperturbed cusp lensing configuration to extract  
information on the matter power spectrum on small scales. 
In this approach, we model our haloes as singular isothermal spheres (SIS).   
A SIS, with density profile $\rho \propto r^{-2}$, is a simple model that is often used   
in lensing because its simplicity permits detailed analytic treatment (e.g.,   
Finch \etal 2002).  
The model has been used to represent mass   
clumps for studies of substructure lensing, after taking into account tidal   
stripping by the parent halo (Metcalf \& Madau 2001; Dalal \& Kochanek   
2002). Again, the simplicity of the SIS makes it attractive for theoretical   
studies: a tool that not only reveals, but also elucidates, some  
interesting general principles.  
For the $10^6 M_{\odot}$ haloes relevant for this work, the   
SIS profile does not differ dramatically from the NFW (Navarro, Frenk, \&   
White 1996) profile inferred from cosmological N-body simulations (Keeton 2003).  
Besides, the SIS model yields {\it conservative} results. Since an NFW halo is  
centrally less centrally concentrated than a SIS halo, it is less efficient as a lens  
and therefore would have to be more massive in order to produce a given  
magnification perturbation.   
Macci\`o \& Miranda (2006) have shown that a SIS model will induce
lensing effects marginally stronger then those caused by an NFW profile   
with concentration parameter $c \sim 55$, corresponding to a mass around $10^6 \Msun$.  
Haloes in a WDM model are expected to be less concentrated due to the top-down structure 
formation scenario (Eke, Navarro \& Steinmetz 2001, Paduroiu \etal 2007 in prep). 
In this case the SIS approximation can possibly overestimate the total effect of WDM perturbers, 
making our lower bound to the WDM particle mass even stronger.

\begin{figure}   
\plotone{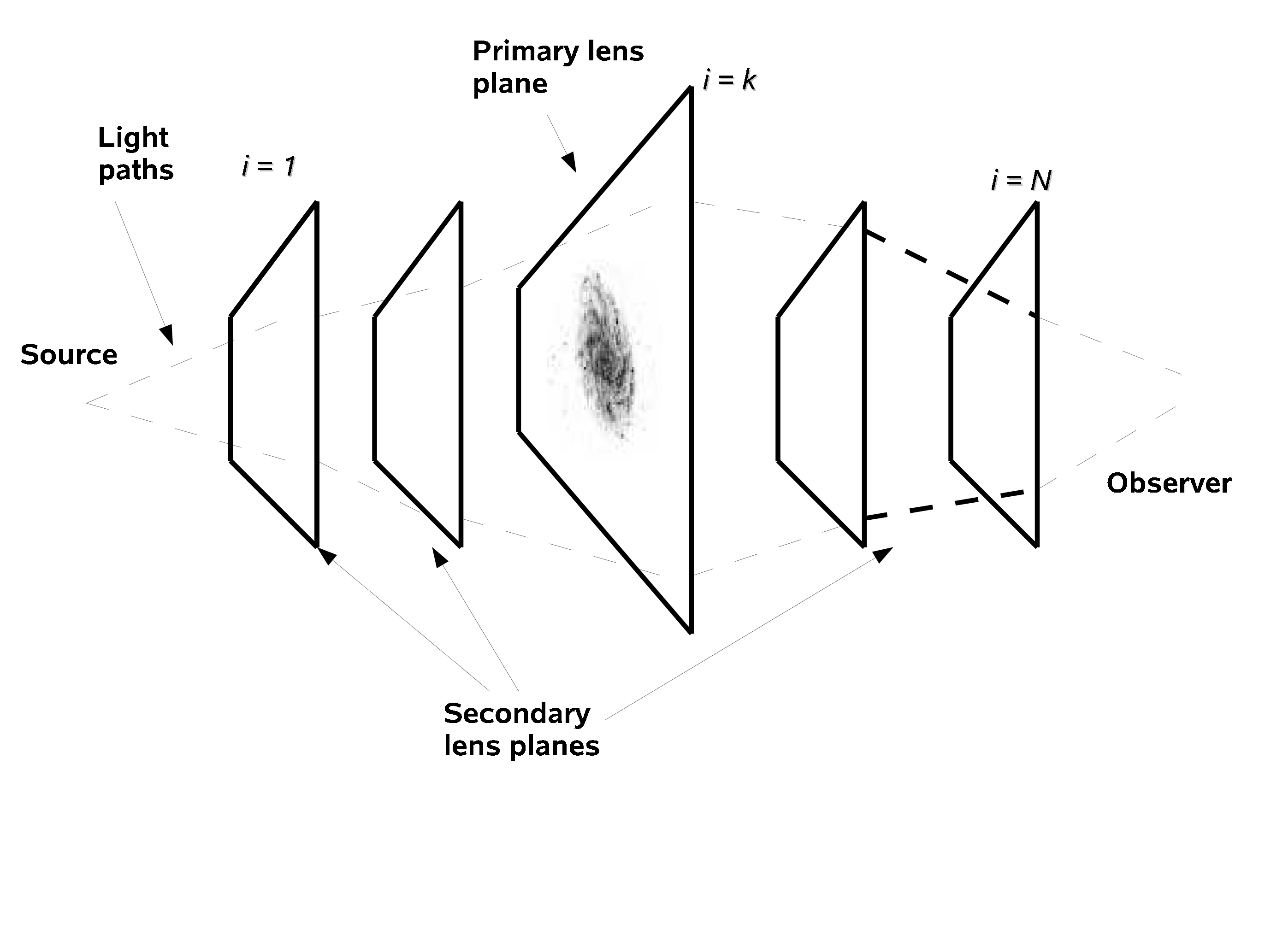}
\caption{A schematic diagram of the type of lensing   
system being considered.  There is one primary lens 
responsible for the multiple images of the source.  In addition,   
there are many secondary lenses (most not shown).  The unperturbed   
light paths are deflected only by the primary lens and with an   
appropriate model for the primary lens will meet on the source   
plane.  If the deflections from secondary lens planes are taken into   
account without changing the primary lens model, the light will follow   
the perturbed light paths (dashed curves). This diagram is not to scale in   
any respect.}   
\label{fig:lensplane2}   
\end{figure}

A SIS halo model is completely characterized by its Einstein radius:   
\begin{equation}   
\theta_{E} = {{4\pi\sigma^2}\over{c^2}} {{D_{LS}}\over{D_S}},   
\end{equation}   
where $\sigma$ is the halo velocity dispersion, and $D_S, D_{LS}$ are the angular  
diameter distances introduced in sec. \ref{sec2los}.  
We adopt a source redshift $z_{s} = 2$. 
  
We filled the portion of Universe along the line of sight with cubes, then the subhaloes   
inside each cube were projected onto the middle plane (see figure \ref{fig:lensplane2}). 
We used a total of 100 different lens   
planes roughly equally distributed in space between the source and the observer.   
This results in $N_1=85$ planes behind the main lens and $N_2=15$ planes in
front of it. The size of the cubes was defined as follows.
  
Two close planes were separated by  $\Delta z_1= (z_{max}-z_l)/N_1$ 
if situated behind the main lensing galaxy, and by $\Delta z_2=(z_{l}-z_{min})/N_2$
for planes in front if it, where $z_{min}=0.01$ and $z_{max}=z_{s}-0.1$.   
  
The size of a comoving volume inside a solid angle $d\Omega$ and a redshift interval $dz$  
is given by (Hogg 1999):  
\begin{equation}  
dV_{\rm C}= D_{\rm H}\,\frac{(1+z)^2\,D_{\rm A}^2}{E(z)}\,d\Omega\,dz  
\end{equation}  
where $D_{\rm A}$ is the angular diameter distance at redshift $z$ and  
$E(z)$ is defined as:  
\begin{equation}  
E(z)\equiv\sqrt{\Omega_{\rm M}\,(1+z)^3+\Omega_k\,(1+z)^2+\Omega_{\Lambda}}  
\label{eq:ez}  
\end{equation}  
with $\Omega_{\rm M}$, $\Omega_k$ and $\Omega_{\Lambda}$ the density  
parameters of matter (cold and warm), curvature, and cosmological constant, respectively.  

We populated each cube with dark matter haloes, whose total number and mass   
distribution was chosen according to the ST mass function at the appropriate redshift  
(see section \ref{sec:haloes}). Halo positions and redshifts (within $\Delta z_{1,2}$)   
were randomly assigned.  
Within a solid angle $d \Omega$ of 3''$\times$3'' squared arcsec,   
the total number of haloes with mass larger than $10^6 \Msun$ comes to 512 for   
the $\Lambda$CDM model adopted in this paper.
This number drops in a consistent way in a warm dark matter scenario, depending on $m_{\nu}$.  
For a WDM particles mass of 10 keV we obtain 238 haloes along the line of sight within the same   
$d \Omega$, and even fewer (156, 135) for a less massive choice for   
$m_{\nu}$ (7.5, 5 keV, see fig \ref{fig:mf}).

Since we are interested in flux anomalies, we consider only cases in   
which we do not have image splitting due to the extra haloes along the line of sight.  
Therefore we do not allow any of those haloes to be closer than twice its Einstein radius   
($\theta_E$) from any images, in order to prevent image splitting (see   
Schneider, Ehlers \& Falco 1992 and references therein). 
{ 
On average only few haloes (~3, for LCDM) 
fail in satisfying this criterium and we tested their removal/inclusion do not affect the final 
$R$ distribution in any way.
Let $\eta$ denote the two-dimensional position of the unperturbed image
with respect to the perturber on the $I$ plane , measured with respect to the intersection 
point of the optical axis with the $I$ plane and $\xi$ the light ray impact parameter on 
the $I^{\prime}$ plane. }
In the absence of image splitting a SIS perturber will affect { the position    
of each image according the following}:     
{ 
\begin{equation}    
\eta = \xi {{D_{I}} \over {D_{I}^{\prime}}} - \alpha({\xi})D_{I^{\prime}I}    
\label{eq:sis}.
\end{equation}   
Introducing the angular coordinates $\eta=D_{I}\theta_{I}$ and
$\xi=D^{\prime}_{I}\theta^{\prime}_{I}$, and given that  $\alpha({\xi})= \theta_E$ for a SIS,}
the equation for the flux becomes
\begin{equation}    
\mu =  {{\theta_{I}^{\prime}} \over {\theta_{I}^{\prime} - \theta_{E}}},    
\label{eq:sis2}
\end{equation}     
where { the quantities with subindex $I$} refer to the (unperturbed) image position with 
respect to the perturber and { so} $D_{I},D_{I^{\prime}},D_{I^{\prime}I}$
are the distances between { observer and the $I$ plane, observer and
  $I^{\prime}$ plane, $I$ plane and $I^{\prime}$ plane, respectively}.
On each single lens plane the total effect on the image magnification
factor $\mu$ is obtained by summing up contributions by each perturber.
In principle one should sum the magnification tensors first and then take the
determinant. The two methods (scalar or matrix sum) do not lead to the same result because $\rm{det} (A+B) 
\ne \rm{det} (A) + \rm{det} (B)$. 
In the case of scalar sum and two SIS perturbers with Einstein radii $\theta_{E,1}$ and $\theta_{E,2}$,
the total magnification depends on the order in which the two lenses act on the 
source: $\mu_{1,2}$ is different from $\mu_{2,1}$. 
The error introduced by a direct sum is of the order of the ratio between the $\mu_{1,2}$ and $\mu_{2,1}$.
This quantity can be directly computed from eqs: \ref{eq:sis} and \ref{eq:sis2} and it is always 
$< \rm max (\theta_{E,1},\theta_{E,2})/\beta$. In our case, due to the low mass of our 
perturbers, the ratio $\theta_{I}/\theta_{E,i}$ is of the order of 200-800, which gives an error 
less than 1\% for the total $\mu$.
There is still a small chance to have a substructure located at a place where 
$\theta_{I} \approx \theta_{E,i}$. We looked for this possibility and it happened only 8 times over 
100.000 substructure position realizations, giving a negligible effect on the final averaged value 
of $R_{cusp}$.

Generally a matter clump will change the positions of the images   
slightly, so if a lens model is chosen to fit the observed image   
positions perfectly it will not do it anymore after the
perturber is added.  To produce a   
perfectly consistent lens model one would have to adjust the main lens   
model for each realization of the intergalactic haloes.  This is   
very computationally expensive and not necessary in practice.   
The shifts in positions are generally small when the masses of the   
secondary lenses are small ($\approx 0.1\arcsec$ for M $\approx 10^{8} M_{\odot}$   
Metcalf 2005a) and, in addition, since the host lens   
model is degenerate it is ambiguous how it should be adjusted to   
correct for the shift.  The goal here is to reproduce all the significant   
characteristics of the effects induced by the observed lens (image configuration, fluxes)   
so that one can determine whether lenses, that look like the observed ones and have the   
observed ratio anomalies, are common in CDM/WDM models.     
For the source, we adopt the point-like approximation.  
The importance of considering the source size lies mainly in the capability
to disentangle different subhaloes mass limits (Chiba \etal 2005, Dobler \& Keeton2006). 
As remarked by Chang \& Refsdal (1979) and many authors afterwords (see Metcalf  
2004 and references therein), the projected size (on the lens plane) of the emitting 
regions of QSOs are expected to be different and this can be used  to remove, 
eventually, lens model degeneracy and improve the sensitivity to substructure properties.
In our cases, the size of the radio emitting region, when projected on the lens plane, is
expected to be affected by structures with masses larger than $10^5 \Msun$ (Metcalf 2005a,b).

In a single realization of our perturbed lens configuration the light coming from the source   
is deflected by $\approx$ 500 haloes (plus the main lens) before reaching the observer.   
Each one of the three images forming the cusp configuration is shifted and amplified,   
giving as a result a modified  $R_{cusp}$ value, different from the original (unperturbed)   
one of $R_{cusp}=0.09$.   
Sometimes, when a massive halo ($M> 10^8 \Msun$) happens to be close to one of the images,   
this image can be strongly deflected, resulting in a breaking of the   
cusp configuration. In the statistical studies presented here these cases are simply   
excluded from the final sample.   
In total we performed 2,000 realizations (with different random seeds for generating   
masses and positions of perturbers) of each  model (CDM/WDM), obtaining 2,000
different final lensing configurations.   
For some of these final configurations (with high $R_{cusp}$ values), we try to fit image  
positions and magnification factors with the {\sc GRAVLENS} code, using a smooth lens model.    
While is relatively simple to reproduce the image geometrical properties, it   
is never possible to get the right flux ratios, with such a simple model.

\section{Results}  
\label{sec:res}

The first part of this section is devoted to presenting the effects of haloes along  
the line of sight (l.o.s.) on the cusp relation in a standard ($\Lambda$)CDM scenario.  
The plots show the probability distribution for the cusp relation value, 
considering 2,000 different realizations of the same model. 
Those realizations share the same total number of  
perturbers, but differ in their masses (randomly drawn from a ST distribution),  
positions (randomly assigned within the lens plane) and redshifts  
(randomly chosen within $\Delta z_{1,2}$). 
 
The cusp relation defined by equation \ref{eq:rcusp} holds when the source is close 
to the cusp. As soon as the source moves away from the cusp, 
deviations from $R_{cusp}=0$ are observed, even for the smooth lens model. 
On the other hand the closer the source is to the cusp, the smaller is the angle  
spanned from the three images. Therefore, in order to take into account  
the position of the source in evaluating the cusp relation, it is better to  
define the anomalous flux ratio as: 
\begin{equation} 
R = {{2 \pi} \over  {\Delta \theta} } R_{cusp}
\label{eq:cusp2} 
\end{equation} 
where $\Delta\theta$ is the opening angle spanned by the two images with 
positive parity defined from the center of the galaxy. 
With this new definition of the cusp relation a set of three images is said to 
violate the cusp relation if $R > 1$. 
This makes the comparison between simulations and observations much more straightforward. 
For this comparison we used the same data presented in Macci\`o \etal (2005). 
There are 5 observed cusp caustic lenses systems ({ summarized in table \ref{tbl:obs}}):  
{\noindent B0712+472 (Jackson \etal 1998),  
B2045+265 (Koopmans \etal 2003), B1422+231 (Patnaik \& Narasimha 2001), 
RXJ1131-1231 (Sluse \etal 2003) and RXJ0911+0551 (Keeton \etal 2003); the 
first three are observed in the radio band, the last two in optical and IR. 
Three of them violate the reduced cusp relation (i.e. $R> 2 \pi / \Delta \theta $).

\begin{table}
\begin{center}
\begin{tabular}{ccccc}
\hline
lens & $\Delta\theta$ & $R_{\rm cusp}$  & { obs. band} \\
\hline
B0712+472 & $79.8^\circ$ & $0.26\pm 0.02$& radio \\
B2045+265 & $35.3^\circ$ & $0.501\pm 0.035$& radio \\
B1422+231 & $74.9^\circ$ & $0.187\pm 0.006$ & radio \\
RXJ1131-1231 & $69.0^\circ$ & $0.355\pm0.015$ & optical/IR \\
RXJ0911+0551 & $69.6^\circ$ & $0.192\pm 0.011$ & optical/IR\\
\hline
\end{tabular}
\caption{\footnotesize The image opening angles and cusp caustic
  parameters for the observed cusp caustic lenses.}
\label{tbl:obs}
\end{center}
\end{table}

Figure \ref{fig:confr} shows the $R$ probability distribution for 
the three possible categories of perturbers. The dotted (red) line shows the effect of 
subhaloes inside the primary lens that can be directly tested by current numerical 
simulations (i.e with masses $>10^7 \Msun$, Macci\`o \etal 2006).  
The short-dashed (cyan) line shows the effect of lower mass subclumps (still inside 
the primary lens) as measured by Macci\`o and Miranda (2006).  
The solid (blue) line shows the effect of the haloes along the line of sight   
considered in this work; here we considered only haloes with $M>5 \times 10^6 \Msun$. 
As already noticed the first two categories of perturbers fail in reproducing  
the high value tail that arises in the observational data around $R=2$. 
On the contrary, the signal coming from haloes along the l.o.s.  
has a probability distribution which remains almost flat in $R$ range 1-2,  
where 2 (out of 5) of the observed systems lay. 
 
Thanks to this pronounced tail at high $R$ value, haloes filling the light cone between the  
source and the observer can easily account for all the observed cusp systems, providing a  
solution to the anomalous flux ratio issue.  
Our results are in fair agreement with those previously obtained by Metcalf (2005b) and 
seem to confirm that a previous result on the same subject obtained by Chen \etal (2003) 
did underestimate the effects of intergalactic structure. 
Chen \etal (2003) used the cross section (or optical-depth) method to calculate the magnification 
probability distribution. This method is mainly valid for rare events and 
this is not the case since, as shown in section \ref{sec:sim}, the number of 
lensing events is of the order of 500. A more detailed and general comparison of the two methods 
can be found in Metcalf (2005b). 
In Metcalf (2005b) the author used an approach similar to ours making a direct  
lensing simulation in order to compute the effects of haloes along the l.o.s., modelling  
them using an NFW density profile.  
Although in his work the author analyzed each observed configuration separately,  
finding slightly different individual $R$ probabilities for different systems,  
the similarity of the results is a good proof {\it a posteriori} that our assumptions 
of SIS parametrization for perturbers and point-source approximation
did not introduce a strong bias in the results.  
 
In the previous analysis we restricted the mass range to haloes more massive than  
$M=5 \times 10^6 \Msun$.  
In figure \ref{fig:mass5-6} the probability distribution for $R$ is shown  
for two different choices of the minimum halo mass: $M>5 \times 10^6 \Msun$  
(solid, blue line) and $M> 10^5 \Msun$.  
In the latter case the total number of structures is around 5,500
and the lensing simulation code slows down considerably.  
A close comparison of the two histograms clearly shows that considering  
less massive haloes does not not improve the results substantially; 
so in the following we will only consider haloes with $M>5 \times 10^6 \Msun$.

In some cases, when the averaging process is restricted to a lower number  
of realizations ($\sim 200$) we found that the observational   
data are reproduced with a high confidence level as shown in figure \ref{fig:good}. 
These results are probably due to effects induced by  single massive 
perturbers close to a particular image: { or a positive image
is highly magnified or a negative one is demagnified (note that in equation \ref{eq:rcusp} we
consider the absolute values for $\mu_{i}$), providing an anomalous $R$}.   
While with a low number of realizations ($\sim 200$) these single events 
contribute significantly to the global $R$, a higher number of 
realizations ($> 10,000$) permits all the images to be affected by massive clumps, 
smoothing the final probability distribution. 
\begin{figure}  
\plotone{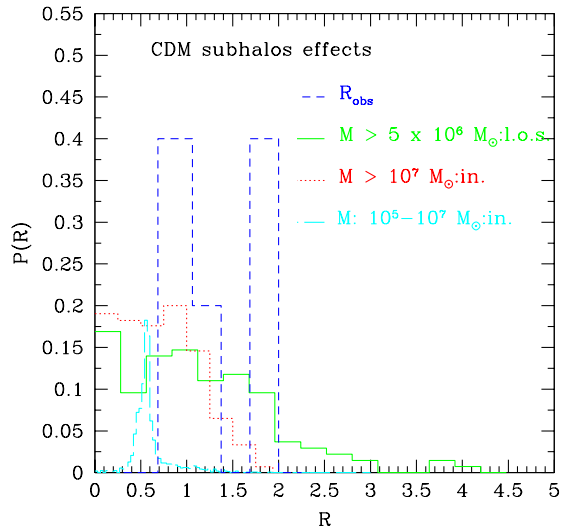}  
\caption{$R$ probability distribution for different categories of (sub)haloes within the CDM 
scenario. The dotted line shows the effect of substructures (with $M>10^7 \Msun $) inside 
the lens galaxy (Macci\`o \etal 2006); the long-dashed line is for less massive subhaloes 
($M=10^5-10^7 \Msun$) still inside the primary lens (Macci\`o \& Miranda 2006). 
The solid line is for the haloes along the line of sight with mass $> 5\times 10^6 \Msun$ 
studied in this work. Observational results are also shown (long dash histogram).
}  
\label{fig:confr}  
\end{figure}  
\begin{figure}  
\plotone{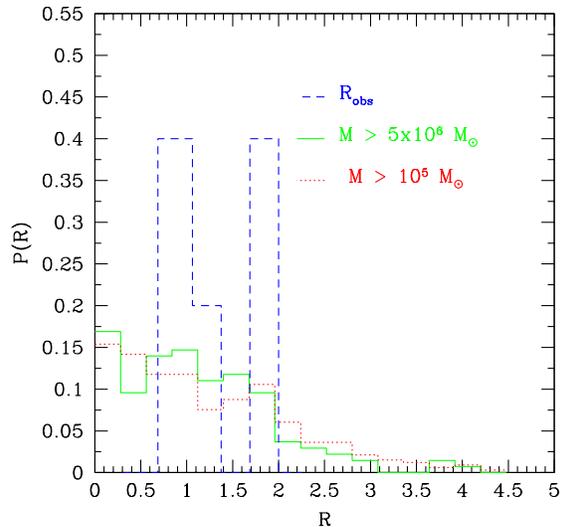}  
\caption{$R$ distribution for haloes along the line of sight for two choices of their minimum mass: 
$M> 10^5$ (dot line) and $M> 5\times10^6 M_{\odot}$ (solid line). The dashed histogram shows 
the observational data.}  
\label{fig:mass5-6}  
\end{figure}  
\begin{figure}  
\plotone{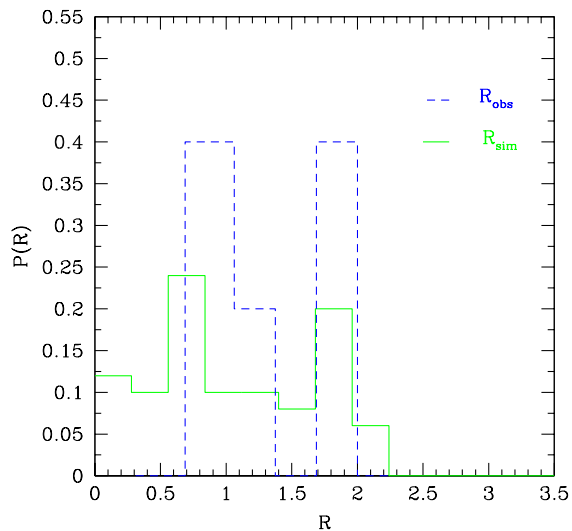}  
\caption{$R$ probability distribution for CDM considering a lower number of realization 
($\approx 200$) in the averaging process (see text). The dashed histogram shows 
the observational data.}
\label{fig:good}  
\end{figure}  
 
The introduction of a WDM particle damps the matter power spectrum on small scales,  
reducing the number of haloes along the l.o.s. 
In figure \ref{fig:wdm_all} we show the probability distribution of $R$
as a function  
of the mass of the WDM candidate. Changing the WDM particle mass from $m_{\nu}=12.5$  
to $m_{\nu} = 7.5$ keV drops the tail at $R=2$ from a 10\% probability to a 1.5\% one.  
For $m_{\nu} = 5$ keV we have a $P(R)$ higher than 5\% only for $R<1.3$. 
In the latter case only 20 haloes are inside the volume sampled by the three images, 
and this model tends to leave the value of $R$ close to the unperturbed one. 
A model with a 10 keV sterile neutrino, if compared to a model with $m_{\nu} = 12.5$ keV, 
gives a slightly lower probability (8\% {\it vs} 10\%) to have a configuration with $R=2$. 
Due to the limited number of observed cusp systems it is hard to disentangle those  
two models, and we think that it is fair to say that $m_{\nu} = 10$ keV is still in  
agreement with the data. 
 
\begin{figure}  
\plotone{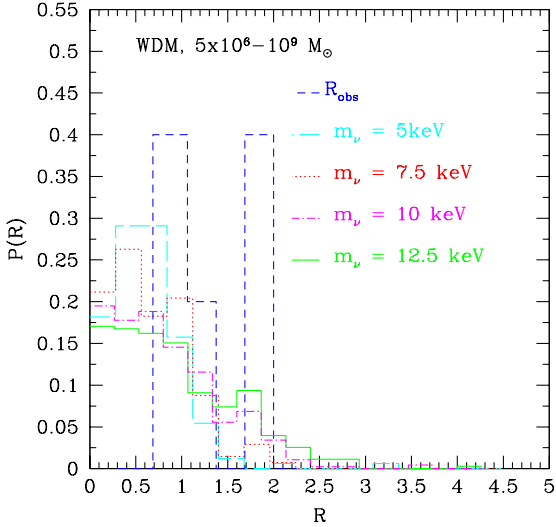}  
\caption{Probability distributions for different warm particle masses:
 $m_{\nu}=5 keV$ (long-dashed line), $m_{\nu}=7.5$ keV (dot line), 
$m_{\nu}=10$ keV (dashed-dot line), $m_{\nu}=12.5$ keV (solid line). 
Dashed line shows the probability distribution of observational data.}  
\label{fig:wdm_all}  
\end{figure}  

Figure \ref{fig:cdmVSwdm} shows the comparison between  
the observational data, the standard ($\Lambda$)CDM model and a WDM model with a  
sterile neutrino mass of 12.5 keV, which is close to the current limit provided by  
Lyman-$\alpha$ + CMB analysis (Seljak \etal 2006).  
In this case in both the warm and cold dark matter scenario, haloes along the line of sight  
can easily account for the two observed cusp systems with $R \approx 2$, offering a  
viable solution to the anomalous flux ratio issue.  
On the contrary a warm dark matter model with less massive particles  
(i.e. with a higher free streaming scale length)  
fails in reproducing the observational data due to the reduced number density of haloes  
along the line of sight. 
 
\begin{figure}  
\plotone{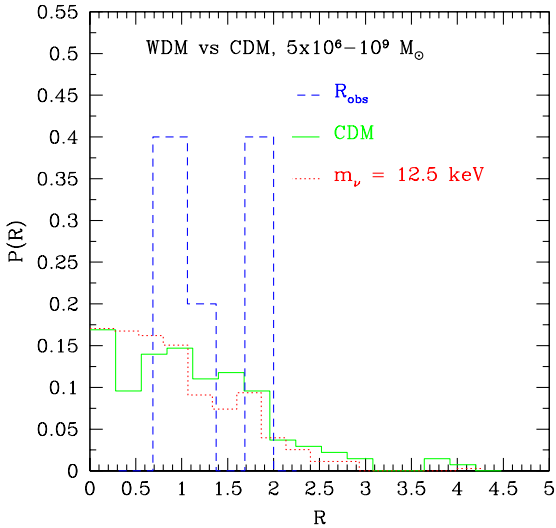}  
\caption{$R$ distribution probability for: observed values (dashed line), CDM haloes more massive 
than $5\times 10^{6} M_{\odot}$ (solid line) and WDM subhaloes with $m_{\nu}=$12.5keV (dotted line).}  
\label{fig:cdmVSwdm}  
\end{figure}  

\section{Discussion and Conclusions}     
\label{sec:disc}   
   
Interest in warm dark matter models has been sporadic over the years,  
although this class of models  could help alleviate several problems on small scales  
that occur with cold dark matter.  
In order to constrain the WDM scenario, precise measurements of the  
matter power spectrum on small scales are needed; for this purpose Lyman-$\alpha$  
forest and CMB data have been extensively used (Seljak \etal 2006, Viel \etal 2006).  
   
In this paper we show that image flux ratios in multiple gravitationally lensed QSOs  
can be modified by haloes along the line of sight in the mass range $10^6-10^7 \Msun$;   
this effect opens a new window to study  the matter power spectrum on small  
scales and provides a new and independent method to constrain the mass of WDM  
candidates $m_{\nu}$.   
   
The observed anomalous flux ratio in lensed QSOs can be explained by adding small  
perturbations to the smooth model used to parametrize the main lenses.  
Those perturbers can be identified with dark matter haloes that happen to be close  
to the images' light paths.  
Recent results based on numerical N-Body (Amara \etal 2006, Rozo \etal 2006) and  
hydrodynamical simulations (Macci\`o \etal 2006) have shown that it is hard to reconcile  
the observed high number of cusp relation violations with the total number of substructures  
inside the primary lens predicted by the $\Lambda$CDM model.  
This is true even when the limited mass  
resolution of numerical simulations is taken into account (Macci\`o and Miranda 2006).   
   
The hierarchical formation scenario predicts that the universe should be filled by a large  
number (more then $10^3$ per $\mpch ^3$) of dark matter haloes with masses  
$M \approx 10^6 \Msun$.  
We employed the Sheth \& Tormen mass function to estimate the expected number of haloes  
in this mass range along the line of sight of lensed QSOs. We found that on
average there are more  
than 500 haloes in between the source and the observer, within a light cone with an aperture  
of 3 arcsec.   
Using direct lensing simulations and a singular isothermal sphere  
approximation we computed the effects of those haloes on an unperturbed  
cusp configuration. We generated more than 10$^4$ different realizations of our  
global (lens + perturbers) lensing system, varying masses, positions, and number of haloes.   
   
We found that on a statistical basis (averaging on different realizations)  
this class of perturbers can modify consistently the  
fluxes of QSO multiple images at a level comparable to the observed one,  
in good agreement with previous studies on this subject (Metcalf 2005a,b).   
In some cases when the averaging process is restricted to a lower number   
of realizations ($\approx 200$, see figure \ref{fig:good}) we found that the observational  
data are reproduced with a high confidence level.  
   
An important result of our study is that the bulk of the signal on QSO fluxes is due to  
haloes in the mass range $10^6-10^7 \Msun$.   
Since the number density of such haloes, and therefore their effect on the cusp relation,  
can be strongly damped by the presence of a WDM candidate, the observed number of anomalous  
flux ratios can be used to constrain the mass of WDM particles.   
   
Adding an exponential cut-off to the transfer function of WDM models we computed the  
number density of small haloes as a function of the mass of the warm  
particles. We show that if WDM is due to a sterile neutrino, then, in models with $m_{\nu}<10$  
keV,  the number of dark haloes along the line of sight is too low to affect in a consistent  
way the fluxes of lensed QSOs, failing to reproduce the observed abundance of systems with  
high $R$ values. This lower limit for the mass of the sterile neutrino is in good agreement  
with results obtained using different methods.   
   
The main limitation of this study is represented by the few observational data that are  
available in the literature. { However, future experiments such as Dune,
  are likely to observe more then 1000 lensed quasars, of which several
  hundreds should be quadruples due to the magnification bias. It will provide
  new lensing systems to be analyzed and thus more tightly constrain the WDM scenario}.

\section{Acknowledgments}      
   
It is a pleasure to thank S. Hansen for enlightening discussion about warm
dark matter, P. Saha for useful hints on the lensing simulations and K.
Blindert for carefully reading the manuscript.
We also thank the referee (HongSheng Zhao) for useful comments that improved the 
presentation of our work and M. Bartelmann and D. Sluse for discussions during the preparation 
of this paper.
M.M. thanks the MPIA Heidelberg for their hospitality while this paper was being completed.
All the numerical simulations were performed on the zBox1 supercomputer (www.zbox1.org)  
at the University of Z\"urich.  M.M. was partially supported by the Swiss National  
Science Foundation.

\end{document}